\documentstyle[twoside,fleqn,espcrc2,epsfig]{article}

\def\beqn{\begin{eqnarray}}
\def\eeqn{\end{eqnarray}}

\title{
Gauge invariant field strength correlators from RG smoothing \\
and color correlations between topological charge 
clusters \thanks{Presented by the first author at Lattice'99, Pisa, Italy.}
}
\author{
E.--M.~Ilgenfritz\address{Institute for Theoretical Physics,
Kanazawa University, Kanazawa 920-1192, Japan}
and 
S.~Thurner\address{Institut f\"ur Kernphysik, 
Technische Universit\"at Wien, A-1040 Wien, Austria}
}
\begin{document}
\begin{abstract}
Using the renormalization group based smoothing technique we have studied
the gauge invariant field strength correlator at $T\ne0$ and $T=0$ in pure 
$SU(2)$ gauge theory. In conjunction with a cluster analysis, the field
strength correlator is used to study correlations between the clusters
in space and color orientation.  
\end{abstract}
\maketitle

\section{Introduction}
\noindent
With the rise of perfect lattice gauge actions, inverse blocking 
\cite{HASEN} has become a 
viable tool in numerical lattice field theory.
A natural requirement for the action is that a classical solution, 
if interpolated in this way, remains classical on all finer lattices. 
Equilibrium configurations are not classical, nevertheless, a  
topological charge can be unambiguously assigned to the interpolating 
field \cite{HASEN} for generic lattice configurations.
This has led to the hope to describe the topological (instanton) 
structure of Monte Carlo (MC) configurations with the help of the 
new method.  
Our variant of RG smoothing \cite{FEURSTEIN} consists of (1) coarse
graining a MC configuration (blocking in order to strip off UV 
fluctuations of scale $a$) and (2) inverse blocking to interpolate it  
by a smooth configuration (SM) on the original lattice. 

RG-smoothing by our method does not lead to classical configurations. 
Clustering of topological charge and action in dominantly (anti)selfdual 
clusters \cite{MONO98} is a property of SM lattice fields 
inherited from the MC vacuum. Smoothed fields preserve the 
string tension, 
its Abelian dominance and contain a topological susceptibility 
of the right magnitude.  In earlier work we were concentrating on the 
monopole content \cite{MONO98,MONO99} of SM configurations in certain gauges, 
relating it to the topological charge distribution and to other gauge-invariant 
signatures. In our present study on SM samples at $T=0$ we 
found the dimensionless ratio $m^{\frac13}/\chi^{\frac14}\approx 1$, remarkably 
independent of $\beta$, in the maximal Abelian and Laplacian gauges 
($m=$ monopole density, $\chi=$ topological susceptibility).
Coming back to the topological 
structure of the $SU(2)$ vacuum, we have 
explored new capabilities of the RG smoothing technique:
(1) to study the gauge invariant field strength correlators at finite 
temperature \cite{FFCORR} and, more recently, at $T=0$, 
(2) to investigate further the clustering of topological charge (started in
\cite{INSTCORR} with a study of correlations in Euclidean and in color space).  
For the perfect action used in our studies see \cite{MONO98}.

\section{Field strength correlator}
\noindent
The gauge invariant field strength correlator \cite{SVM},
\beqn
{\cal D}_{\mu\rho,\nu\sigma}(x_1-x_2) =  
 \langle 0|  \mbox{\rm tr} \{  G_{\mu\rho}(x_1) S(x_1,x_2)  \nonumber \\
            G_{\nu\sigma}(x_2) S^{\dagger}(x_1,x_2)\} |0 \rangle  
\eeqn
describes the non-perturbative vacuum structure in a model-independent way. 
Decomposing it into two basic structure functions, 
\beqn
{\cal D}_{\mu\rho,\nu\sigma}(x)  =  
 \left(\delta_{\mu\nu}\delta_{\rho\sigma}
 -\delta_{\mu\sigma}\delta_{\rho\nu}\right)
\left({\cal D}(x^2) +{\cal D}_1(x^2)\right) \nonumber \\
+ \left(x_\mu x_\nu \delta_{\rho\sigma} + x_\rho x_\sigma \delta_{\mu\nu}
- ... \right)
\frac{\partial {\cal D}_1(x^2)}{\partial x^2} , \nonumber
\eeqn
exposes a confining part ${\cal D}$, while ${\cal D}_1<<{\cal D}$ is
related to the dominance of (anti)selfdual 
local excitations \cite{MARTEM} 
contributing to it. Using smoothing (instead of cooling \cite{DIGIAC}) 
we want to study eventual effects of smoothing as a noise reduction method
on the topological content. 
The ``Schwinger line'' $S(x_1,x_2)$,
a path dependent transporter,
requires (for distances not on-axis on the lattice) to perform a 
random sampling of paths of shortest length.
We have measured at finite $T$ electric and magnetic structure functions 
${\cal D}_L^E=\left({\cal D}^E+{\cal D}_1^E+{\bf x}^2
\frac{\partial {\cal D}_1^E}{\partial {\bf x}^2}\right)$,
${\cal D}_T^E=\left({\cal D}^E+{\cal D}_1^E\right)$,
${\cal D}_L^B=\left({\cal D}^B+{\cal D}_1^B\right)$ and
${\cal D}_T^B=\left({\cal D}^B+{\cal D}_1^B+{\bf x}^2
\frac{\partial {\cal D}_1^B}{\partial {\bf x}^2}\right)$
at space-like distances for samples of $2000$ smoothed $12^3\times4$ 
configurations at $\beta=1.4$, $1.5$, $1.55$, $1.6$, $1.7$ and $1.8$ 
(a temperature interval $T/T_c=0.62$, $0.86$, $1.03$, $1.25$, $1.97$, $\approx4$).  
The lattice spacing $a(\beta)$ has been calibrated using the zero
temperature $SU(2)$ string tension by comparison with the monopole 
Creutz ratios which reach a plateau early on a $12^4$ lattice 
(except for $\beta=1.8$).
For $T=0$ the same functions have been studied on $12^4$ SM configurations 
at $\beta=1.54$ ($a=0.17$ fm) corresponding to the deconfinement $\beta_c$ 
found on $12^3\times4$ \cite{MONO98} for this action. 

In the confinement phase, the finite temperature correlators have an
almost perfect electric-magnetic symmetry, even on our unsymmetric lattice. 
The signal of the deconfining transition is the breakdown of ${\cal D}_L^E$ 
(Fig. 1).
\begin{figure}[t]
%\vspace{-6mm}
\begin{minipage}{7.5cm}
\begin{center}
\epsfig{file=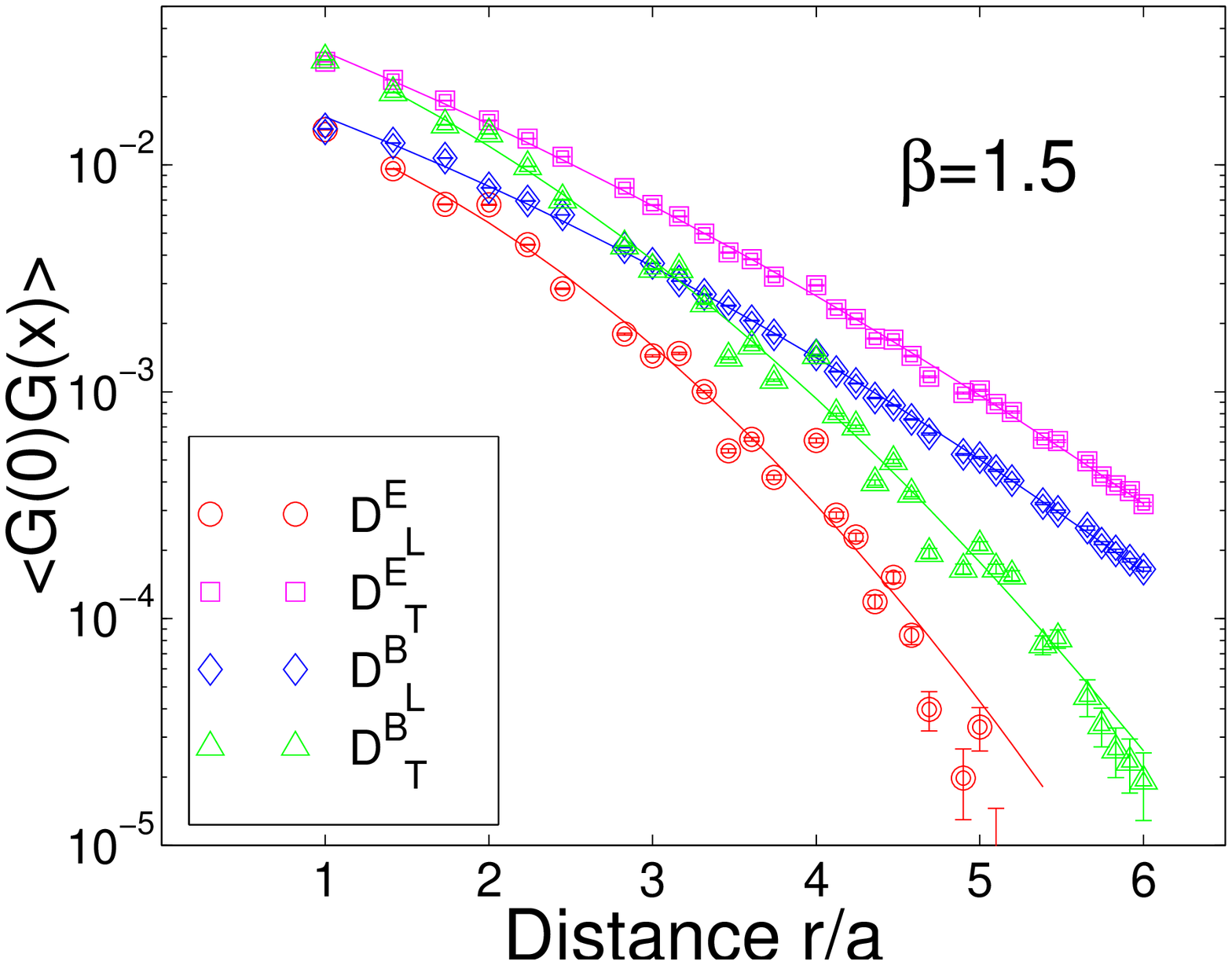,width=3.6cm,height=3.0cm}
\epsfig{file=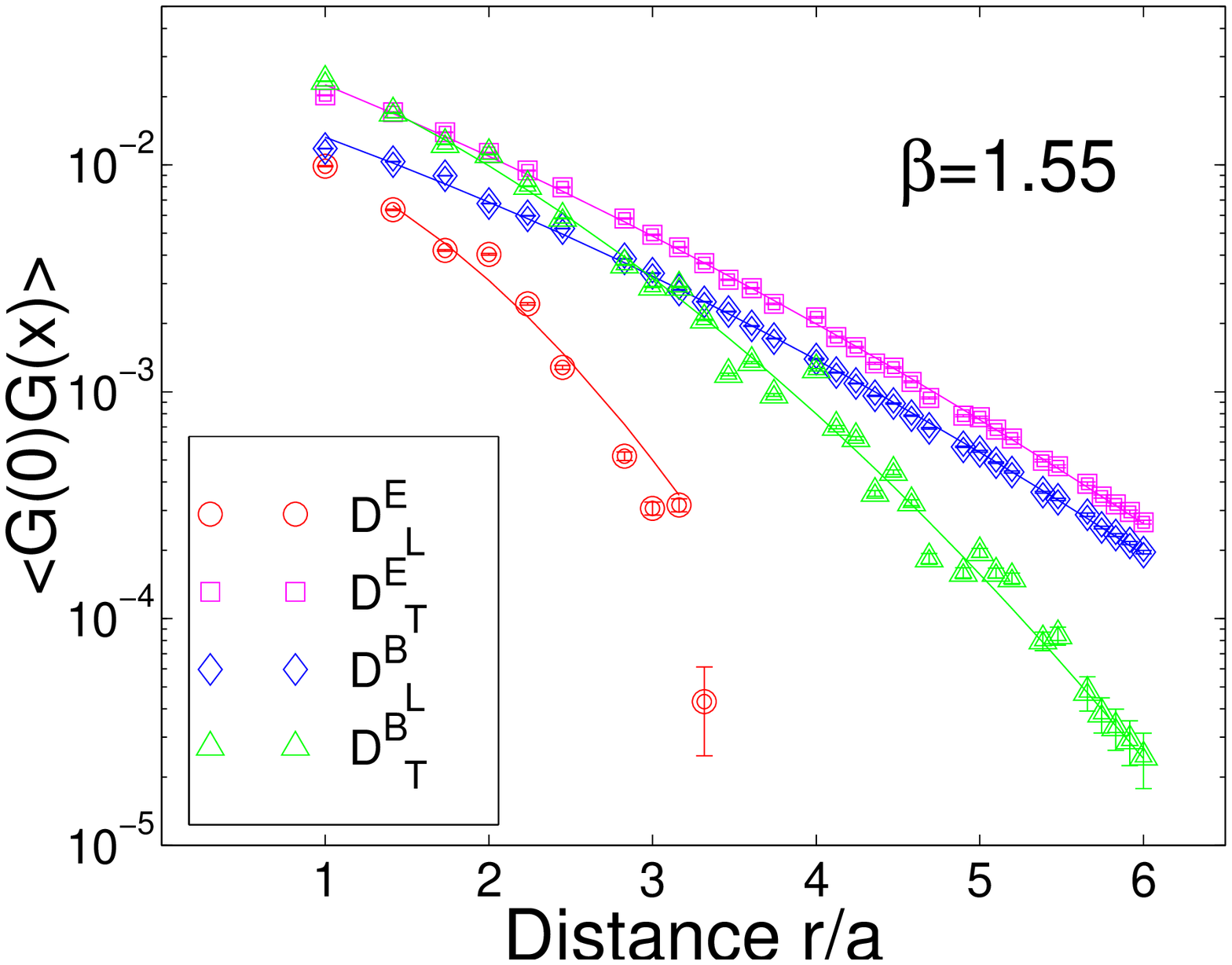,width=3.6cm,height=3.0cm} \\
(a) \hspace{3.0cm} (b)\\
\end{center}
\end{minipage}
\vspace{-8mm}
 \caption{\small Electric and magnetic correlators 
(a) below, (b) above the deconfinement transition.}
\label{raw data}
\vspace{-8mm}
\end{figure}
%%%%  was ist electric/magnetic ?  
A fit with exponential structure functions
${\cal D}^{E,B}$ and ${\cal D}_1^{E,B}$
doesn't give a satisfactory description for all $T$. 
In \cite{FFCORR}
we solve the equations defining ${\cal D}_L^{E,B}$ and ${\cal D}_T^{E,B}$.
This leads to condensates and {\it integrated} correlation lengths 
given in Table 1.
\begin{table}[t]
\label{tab:expfit}
% \vspace{0.2cm}
\caption{Fit of condensates and integrated correlation lengths.}
\begin{center}
\footnotesize
\begin{tabular}{  l c c c c }
\hline
   {$\beta$} 
 & ${\cal D}^E(0)+$ 
 & ${\cal D}^B(0)+$   
 & $\xi_{int}^E$ 
 & $\xi_{int}^B$  \\
 
 & ${\cal D}_1^E(0)$ & ${\cal D}_1^B(0)$ &  &  \\
 &  [GeV$^4$]        & [GeV$^4$]         & [fm]  &  [fm] \\
\hline
1.40  &$0.010(1)$    &$0.010(1)$    &$0.32(1) $  &$0.34(2)$ \\
1.50  &$0.029(2)$    &$0.028(3)$    &$0.24(1) $  &$0.27(1)$ \\
1.55  &$0.044(7)$    &$0.047(3)$    &$0.16(2) $  &$0.24(1)$ \\
1.60  &$0.066(8)$    &$0.083(12)$   &$0.15(3) $  &$0.23(2)$ \\
1.70  &$0.312(20)$   &$0.312(81)$   &$0.07(2) $  &$0.13(2)$ \\
%1.80  &$5.082(123)$  &$6.962(232)$  &$0.03(2) $  &$0.06(1)$ \\
\hline
\end{tabular}
\vspace{-12mm}
\end{center}
\end{table}
For $T=0$ we find $\xi^E$ and $\xi^B$ 
compatible with
the lowest temperature there. 
The values of the correlation lengths 
are in the expected ballpark. 
In terms of ${\cal D}$ and ${\cal D}_1$ the transition appears as the result
of continuous changes.
The degree of (anti)selfduality begins to
decrease already below the deconfining transition
\cite{FFCORR}.
For a better understanding a detailed comparison with the cooling method
(presently not available for $SU(2)$) is under way.
%%%%  Conclusion ?
%ST Da gibts einfach keine in Bezug auf cooling. Muessten wir  
  % einfach noch machen
  
\section{Correlations of  topological clusters}
\noindent
The uncorrelated instanton liquid provides a good description
of the QCD ground state. There are no indications 
from the field strength correlator at $T=0$ \cite{MARTEM}  
for strong correlations between (anti)instantons ($I$ and $A$).
At higher temperatures interactions are expected to be important  
in the instanton ensemble. Contrary to fermionic interactions,
Ans\"atze for gluonic interactions rely on specific
$II$ and $IA$ superpositions.
Analyzing the clustered SM configurations we hoped to learn about
instantons in the real Yang Mills vacuum. Instead of supporting the 
instanton liquid picture our results are more suggestive for a different 
interpretation of the clusters exposed by smoothing.

The charge density has been measured by means of L\"uscher's charge
which is inexpensive for SM configurations. 
We have defined topological clusters using a $4-D$ site percolation 
algorithm for assigning neighboring (same sign) lattice sites, marked 
to have $|q(x)|>q_{th}$ (threshold), to the same cluster. 
Clusters are characterized by a center (identified by $q_{max}$), 
a cluster volume $V_{cl}$ and cluster charge $Q_{cl}$. 
Eventual color correlations between different clusters can be studied
by the normalized {\it cluster overlap} for a pair of clusters,
\beqn
{\cal O} =
  \frac{ \langle {\mathrm tr}\left(G_{\mu\nu}(1)~S(1,2)
		       ~G_{\mu\nu}(2)~S(2,1)\right)
			\rangle_{\mathrm{pathes}}
		   }
{ ({\mathrm tr}\left(G_{\rho\sigma }(1)^2\right))^{\frac12}
  ~({\mathrm tr}\left(G_{\tau\lambda}(2)^2\right))^{\frac12} }
\eeqn
with the field strength correlator between the centers involved.
Correlators of specified components would give, for instanton pairs,
access to the relative color orientation matrix $U_{rel}$.
The overlap ${\cal O}$ (2) can be identified with the adjoint trace
$\left(({\mathrm tr}U_{rel})^2-1\right)/3$. For random orientations 
the average of ${\cal O}$ over all pairs would vanish. Plotting the
average for different pairs as a function of the distance shows 
that the clusters are {\it not uncorrelated} (Fig. 2). This average is
sensitive to deviations of the histogram of ${\mathrm tr}U_{rel}$ from
the Haar measure. Maximizing the overlap by a gauge rotation one can find 
$U_{rel}$ 
for each pair. This method can be applied in all instanton
finding  studies.
\begin{figure}[t]
%\vspace{-12mm}
\begin{center}
\epsfig{file=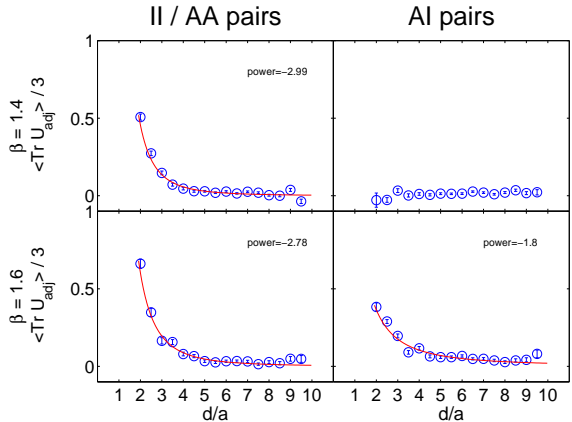,width=7.5cm,height=5.7cm}
\vspace{-10mm}
\caption{\small Average cluster overlap ${\cal O}$ vs. distance, in
confinement (top) and deconfinement (bottom) for same sign and opposite 
sign clusters.}
\label{type}
\vspace{-7mm}
\end{center}
\end{figure}
Our analysis has shown for $II$ pairs strong {\it aligning} correlations 
in the confined phase (decreasing with distance) but none for $IA$ 
pairs and {\it aligning} correlations {\it similar} for both types of pairs
in the deconfined phase. A recent $T=0$ study confirms strong $II$ correlations
and rather weak ones for $IA$.
These results are difficult to reconcile with any instanton
picture.

Therefore we have carefully studied the influence of the cutoff   
$q_{th}$ (over an interval of space filling fraction between 1 and 10 \%) 
on the cluster composition. The cluster multiplicity goes through a maximum
before a cluster percolation transition towards 
huge and multiply charged clusters is observed.
With a cutoff keeping the cluster multiplicity below the maximum we get
clustering properties being rather 
cutoff independent (apart from distance correlations between 
centers) with an almost Poissonian multiplicity distribution. 
For instance, on the $12^4$ lattice at 
$\beta=1.54$ ($\langle Q^2 \rangle \approx 11$) 
we find for $q_{th}=0.065$ an average number of clusters of $\approx 35$.
A strong correlation exists relating charge and volume,
$|Q_{cl}| \approx 0.01 V_{cl}$ (with almost all $|Q_{cl}| < 0.5$).

\section{ Conclusion}
\noindent
If these clusters identified on SM configurations by L\"uscher's charge
could be interpreted as subclusters of instantons (``instanton quarks''=
half-instantons for $SU(2)$) this would explain the unexpected pattern
of color correlations. Equal sign clusters would be bound into instantons
(dipoles) in the confinement phase. They could dissociate in the deconfined 
phase with opposite charge correlations becoming of equal importance. 
Further investigations are worthwhile, relating this to the behavior of
the field strength correlators above.
If this picture can be consolidated
this would be a step beyond the present search for instanton structure by 
various cooling techniques \cite{TEPER}.

\end{document}